\documentclass[widetext,showpacs,preprintnumbers,superscriptaddress,nofootinbib]{revtex4}

\usepackage{graphicx}
\usepackage{amssymb}
\usepackage{amsmath}
\usepackage{dcolumn}
\usepackage{bm}
\usepackage{natbib}
\usepackage{multirow}
\usepackage{subfigure}

\begin{document}

\preprint{ICRR-Report-659-2013-8,\ IPMU13-0182,\ YITP-13-87}

\title{On the estimation of gravitational wave spectrum from cosmic domain walls}

\author{Takashi Hiramatsu}
\email{hiramatz@yukawa.kyoto-u.ac.jp}
\affiliation{Yukawa Institute for Theoretical Physics, Kyoto University,\\
Kitashirakawa Oiwake-cho, Sakyo-ku, Kyoto 606-8502, Japan}

\author{Masahiro Kawasaki}
\email{kawasaki@icrr.u-tokyo.ac.jp}
\affiliation{Institute for Cosmic Ray Research, The University of Tokyo,\\
5-1-5 Kashiwa-no-ha, Kashiwa City, Chiba 277-8582, Japan} 
\affiliation{Kavli Institute for the Physics and Mathematics of the Universe (WPI), Todai Institutes for Advanced Study, The University of Tokyo,\\
5-1-5 Kashiwa-no-ha, Kashiwa City, Chiba 277-8582, Japan}

\author{Ken'ichi Saikawa}
\email{saikawa@th.phys.titech.ac.jp}
\affiliation{Department of Physics, Tokyo Institute of Technology,
2-12-1 Ookayama, Meguro-ku, Tokyo 152-8551, Japan}
 
\date{\today}

\begin{abstract}
We revisit the production of gravitational waves from unstable domain walls analyzing their spectrum by the use of field theoretic lattice simulations 
with grid size $1024^3$, which is larger than the previous study.
We have recognized that there exists an error in the code used in the previous study, and the correction of the error
leads to the suppression of the spectrum of gravitational waves at high frequencies.
The peak of the spectrum is located at the scale corresponding to the Hubble radius at the time of the decay of domain walls, and its amplitude
is consistent with the naive estimation based on the quadrupole formula.
Using the numerical results, the magnitude and the peak frequency of gravitational waves at the present time are estimated.
It is shown that for some choices of parameters the signal of gravitational waves is strong enough to be probed in
the future gravitational wave experiments.
\end{abstract}

\pacs{98.80.Cq,\ 04.30.Db}

\maketitle

\begin{widetext}
\tableofcontents\vspace{5mm}
\end{widetext}

\section{\label{sec1}Introduction}
Gravitational wave~\cite{Maggiore:1900zz} is one of the most promising observational probes of the physics of the early universe.
It has been considered that there exist many possibilities to
generate observable signatures of gravitational waves in the early universe~\cite{Maggiore:1999vm,Binetruy:2012ze}.
These include quantum fluctuations during inflation~\cite{Grishchuk:1974ny},
the first order phase transition~\cite{Witten:1984rs}, cosmic strings~\cite{Vilenkin:1981bx},
non-perturbative decay of the inflaton (preheating)~\cite{Khlebnikov:1997di,Easther:2006gt}, and so on.
It is expected that the observation of such signals of gravitational waves significantly improves our understandings of fundamental physics.
These days a number of experimental searches including ground-based~\cite{Abramovici:1992ah,Acernese:2008zzf,Willke:2002bs,Somiya:2011np,Sathyaprakash:2012jk}
and space-borne~\cite{Seto:2001qf,AmaroSeoane:2012km} observations are running or planned.
Accordingly, it is important to understand the nature of sources that are relevant to observations.

In this paper, we study domain walls as a source of primordial gravitational waves.
Domain walls are sheet-like objects which are formed in the early universe when a discrete symmetry is spontaneously broken [see e.g.~\cite{2000csot.book.....V}].
It is known that the existence of stable domain walls conflicts with the standard cosmology,
since their energy density tends to dominate over the total energy density of the universe~\cite{Zeldovich:1974uw}.
However, it remains a possibility to consider the case where domain walls are unstable,
which annihilate at sufficiently early times in order not to affect the standard cosmological history~\cite{Vilenkin:1981zs,Gelmini:1988sf,Coulson:1995nv,Larsson:1996sp}.
Such unstable but long-lived domain walls would be another source of
the stochastic gravitational wave background observed today~\cite{Gleiser:1998na,Hiramatsu:2010yz,Kawasaki:2011vv}.

In the literature, it is discussed that several particle physics models predict the formation of unstable domain walls and the signature of gravitational waves produced from them.
For instance, the spontaneous breaking of the discrete R-symmetry in supersymmetric theories would lead to the formation of domain walls~\cite{Dine:2010eb},
and gravitational waves produced from them can be regarded as a probe of the gravitino mass~\cite{Takahashi:2008mu}.
Domain walls are also formed in the thermal inflationary scenario~\cite{Moroi:2011be}, which also predicts signatures relevant to observations.
In addition, the production of gravitational waves from domain walls is discussed
in the context of the next-to-minimal supersymmetric standard model (NMSSM) in~\cite{Hamaguchi:2011nm},
the extension of the standard model with right-handed neutrinos in~\cite{Ishida:2013mva}, 
and the axion models in~\cite{Hiramatsu:2010yn,Hiramatsu:2012sc}.
Predictions of these models rely on the form of the spectrum of gravitational waves, which necessitates further investigations.

The spectrum of gravitational waves from domain walls is computed by using the numerical simulation of the classical scalar field
in Refs.~\cite{Hiramatsu:2010yz,Kawasaki:2011vv}.
In these previous studies, it was found that the spectrum becomes almost flat extending intermediate frequencies between the scale corresponding to the horizon size
at the decay time of domain walls and that corresponding to the width of them.
However, the shape of the spectrum and its dependence on some theoretical parameters are not fully understood,
due to the lack of the dynamical range of the numerical simulations.
In this work, we check the previous results~\cite{Hiramatsu:2010yz,Kawasaki:2011vv} by performing high-resolution simulations with $1024^3$ grids.
The large dynamical range of the current simulation enables us to investigate the dependence on the parameter controlling the width of domain walls,
which was not investigated in the previous studies.
Furthermore, in updating the numerical code we recognized that the code used in the previous studies~\cite{Hiramatsu:2010yz,Kawasaki:2011vv}
contained an error in the evaluation of the transverse-traceless projection of the stress-energy tensor.
In this paper we correct this error and discuss its influence on the numerical results.
Correcting the error does not modify the result on the estimation of the overall amplitude of gravitational waves,
but the amplitude becomes suppressed at high frequencies with a slope $\sim k^{-1}$, rather than the flat spectrum observed in the previous results.

The rest of this paper is organized as follows.
In Sec.~\ref{sec2}, we briefly review the evolution of the network of unstable domain walls.
After that, in Sec.~\ref{sec3} we describe the analysis method to calculate the evolution of domain walls and gravitational radiations from them.
Results of the numerical study are presented in Sec.~\ref{sec4}.
In Sec.~\ref{sec5}, we estimate the amplitude and frequency of gravitational waves observed today by using the numerical results.
Finally, we conclude in Sec.~\ref{sec6}.
Details of the error in the numerical code are described in Appendix~\ref{secA}.

\section{\label{sec2}Domain wall network evolution}
In this paper, we work in the spatially flat Friedmann-Robertson-Walker (FRW) universe,
\begin{equation}
ds^2 = dt^2 - a^2(t)[dx^2+dy^2+dz^2], \nonumber
\end{equation}
where $a(t)$ is the scale factor of the universe.
Then we consider the model of the real scaler field $\phi$, whose Lagrangian density is given by
\begin{equation}
\mathcal{L} = \frac{1}{2}\partial_{\mu}\phi\partial^{\mu}\phi - V(\phi). \label{eq2-1}
\end{equation}
For the scalar potential $V(\phi)$, we adopt the following double well potential
\begin{equation}
V(\phi) = \frac{\lambda}{4}(\phi^2-\eta^2)^2. \label{eq2-2}
\end{equation}
In the early universe with a finite temperature $T$, the correction term $\lambda T^2\phi^2/8$ would be added to this potential,
and the discrete $Z_2$ symmetry ($\phi\to-\phi$) is restored.
When the temperature is cooled  below the critical value $T_c = 2\eta$, this $Z_2$ symmetry is spontaneously broken
to form domain walls.

Once the domain walls are formed, their curvature radius is rapidly homogenized due to the surface tension of them.
Eventually they evolve toward the ``scaling regime" where the typical length scales of the network of them,
such as the curvature radius and the distance between neighboring walls, become comparable to the Hubble radius $H^{-1}\sim t$.
In other words, there always exists one (or a few) domain wall(s) inside the horizon in the scaling regime.
The appearance of this scaling property is confirmed by various numerical and
analytical studies~\cite{Press:1989yh,Garagounis:2002kt,Oliveira:2004he}.

Since the typical size of domain walls is given by the Hubble radius
in the scaling regime, the energy density of domain walls evolves as
\begin{equation}
\rho_{\rm wall} \sim \frac{H^{-2}\sigma_{\rm wall}}{H^{-3}} \sim \sigma_{\rm wall}/t, \label{eq2-3}
\end{equation}
where $\sigma_{\rm wall} = 2\sqrt{2\lambda}\eta^3/3$ is the surface mass density of domain walls.
Hence it would be useful to introduce the scaling parameter (or area parameter)~\cite{Hiramatsu:2012gg,Hiramatsu:2012sc} given by
\begin{equation}
{\cal A} \equiv \frac{\rho_{\rm wall}}{\sigma_{\rm wall}}t. \label{eq2-4}
\end{equation}
We expect that the quantity ${\cal A}$ remains constant during the scaling regime.
This property will be checked by numerical simulations in Sec.~\ref{sec4}.

In the scaling regime, domain walls frequently interact with each other, changing their configuration or collapsing into the closed walls,
to reduce their energy and maintain the scaling property~\eqref{eq2-3}.
During such a process, a fraction of the energy of domain walls is released as gravitational waves.
The magnitude of gravitational waves radiated from them can be estimated by using the quadrupole formula~\cite{Hiramatsu:2012sc}.
With the assumption that the typical time scale of the gravitational radiation is comparable to the Hubble time,
the power of gravitational radiation is given by $P\sim G\dddot{Q}_{ij}\dddot{Q}_{ij}\sim M_{\rm wall}^2/t^2$, where
$G$ is the Newton's constant, $Q_{ij}\sim M_{\rm wall}t^2$ is the quadrupole moment of the domain walls,
and $M_{\rm wall}\sim \sigma_{\rm wall}{\cal A}t^2$ is the mass energy of them.
Therefore, the energy density of gravitational waves is estimated as
\begin{equation}
\rho_{\rm gw} \sim \frac{Pt}{t^3} \sim G\mathcal{A}^2\sigma^2_{\rm wall}. \label{eq2-5}
\end{equation}
Note that $\rho_{\rm gw}$ does not depend on the cosmic time $t$.
Therefore, we introduce the efficiency parameter~\cite{Hiramatsu:2012sc}
\begin{equation}
\epsilon_{\rm gw} \equiv \frac{\rho_{\rm gw}}{G\mathcal{A}^2\sigma_{\rm wall}^2}, \label{eq2-6}
\end{equation}
which is expected to be constant in the scaling regime.
In Sec.~\ref{sec4}, we will 
see that the value of $\epsilon_{\rm gw}$ becomes almost constant at late times of the numerical simulations.

The existence of the long-lived domain wall networks is problematic, since the energy density of them ($\rho_{\rm wall}\propto t^{-1}$)
decreases slower than that of matters and radiations. If they are stable, they eventually overclose the universe, which might conflict with the
standard cosmological scenario.
In order to avoid this problem, they must decay at sufficiently early times.
The decay of domain walls can be promoted by introducing a so-called ``bias" term $\delta V$ in the potential, which explicitly breaks the discrete symmetry.
As in the previous studies~\cite{Hiramatsu:2010yz,Kawasaki:2011vv}, we use the following term to model this effect
\begin{equation}
\delta V = \epsilon\eta\phi\left(\frac{1}{3}\phi^2-\eta^2\right), \label{eq2-7}
\end{equation}
and treat $\epsilon$ as a free parameter. This term lifts the degenerate vacua
and induces the difference in the energy density between them $V(\eta)-V(-\eta) = 4\epsilon\eta^4/3$.
This difference in the energy density affects on the wall as a volume pressure, $p_V \simeq 4\epsilon\eta^4/3$.
The annihilation of the domain wall networks occurs when this volume pressure becomes comparable to the
surface tension of the wall $p_T\simeq \sigma_{\rm wall}/R$, where $R\sim H^{-1}$ is the curvature radius of them.
From the condition $p_T\simeq p_V$, we can estimate the decay time of walls as
\begin{equation}
t_{\rm dec} \simeq \frac{1}{2}\sqrt{\frac{\lambda}{2}}(\epsilon\eta)^{-1}, \label{eq2-8}
\end{equation}
where we assumed the radiation dominated universe to use $H^{-1}=2t$.

The magnitude of the parameter $\epsilon$ is assumed to be sufficiently small such that it breaks the discrete symmetry
only approximately. 
In order that domain walls with infinite size are formed, it must satisfy $\epsilon < 0.15\lambda$~\cite{Hiramatsu:2010yz}.
Furthermore, we can obtain the following lower bound on $\epsilon$ by requiring that domain walls disappear before they overclose the universe,
\begin{equation}
\epsilon > \frac{16\pi}{9}\lambda\left(\frac{\eta}{M_P}\right)^2, \label{eq2-9}
\end{equation}
where $M_P$ is the Planck mass.
The possible scenarios with biased domain walls and constraints on $\epsilon$ are discussed in Refs.~\cite{Gelmini:1988sf,Hiramatsu:2010yz}.

Although the formulae~\eqref{eq2-8} and~\eqref{eq2-9} are derived from the bias term of the form~\eqref{eq2-7},
other terms which explicitly violate the discrete symmetry (such as $\delta V =\epsilon\eta^3\phi$) induce similar effects.
If we use the bias term whose form is different from Eq.~\eqref{eq2-7}, the numerical coefficients of Eqs.~\eqref{eq2-8} and~\eqref{eq2-9}
would be modified, which just affects the estimation of the decay time of domain walls by a factor of $\mathcal{O}(1)$.

It should be noted that we perform the simulation without including the explicit symmetry breaking term~\eqref{eq2-7}, and do not
simulate the collapse of domain walls.
Instead, we determine the numerical coefficients such as $\mathcal{A}$ and $\epsilon_{\rm gw}$, which are defined in the scaling regime,
from the results of the simulations.
Then, we estimate the present density and frequency of gravitational waves in Sec.~\ref{sec5} with the assumption that domain walls suddenly disappeared
at the time given by Eq.~\eqref{eq2-8}.
Therefore, the dependence on the parameter $\epsilon$ appears in the analytic formulae presented in Sec.~\ref{sec5}
through the relation~\eqref{eq2-8}.

There are two reasons why we drop the bias term~\eqref{eq2-7} in the present numerical study.
Firstly, the inclusion of the bias term leads to a subtlety in the estimation of parameters characterizing the scaling regime.
The numerical simulations of domain wall networks with the bias term were performed in Refs.~\cite{Larsson:1996sp,Hiramatsu:2010yz}.
In these works, it was confirmed that the area of domain walls within the simulation box rapidly falls off in the time scale estimated by
Eq.~\eqref{eq2-8}. However, in such kind of the simulation, it is difficult to reproduce the scaling regime before the onset of the decay of domain
walls due to the limitation of the dynamical range, and the numerical factors such as $\mathcal{A}$ and $\epsilon_{\rm gw}$
are not determined unambiguously.
The second reason is that the effect of the bias term might be negligible as long as we are interested in the form of the spectrum
relevant to the observations.
In Ref.~\cite{Hiramatsu:2010yz} it was found that the decay of domain walls
leads to make a peak-like feature in the high frequency region of the gravitational wave spectrum,
which can be interpreted as the fragmentation of the false vacuum regions into small pieces.
As we will see in the following sections, such a feature at the high frequency is irrelevant to the observations, while
it is possible to observe the low frequency region comparable to the Hubble parameter at the decay time.
With the expectation that the spectrum of interest in observations (i.e. the spectrum at low frequencies)
preserves the form produced in the scaling regime, in this work we estimate the spectrum at the present time
by extrapolating the result of the numerical simulation obtained in the scaling regime, and ignore the effect of the bias term.

\section{\label{sec3}Simulation techniques}
In this section, we describe some technical aspects of the numerical study to calculate the quantities such as
the scaling parameter and the spectrum of gravitational waves.
We solve the equation of motion of the scalar field in the FRW background on the three dimensional lattice with the periodic boundary condition,
\begin{equation}
\ddot{\phi} + 3H\dot{\phi}-\frac{\nabla^2}{a^2}\phi + \frac{dV}{d\phi} = 0, \label{eq3-1}
\end{equation}
with the potential given by Eq.~\eqref{eq2-2}.
The spatial derivative is computed by using the fourth order finite difference method,
and the time evolution is solved by using the fourth order symplectic integration scheme~\cite{Yoshida:1990zz}.
Note that we are not using the PRS algorithm~\cite{Press:1989yh,Sousa:2010zza}, in which the equation of motion is modified from
Eq.~\eqref{eq3-1} such that the width of domain walls is fixed in the comoving simulation box.
This algorithm enables us to improve the dynamical range of the simulation since we do not have to
care about the resolution of the width of domain walls [see discussions below Eq.~\eqref{eq3-19}].
However, if we use this algorithm we incorrectly estimate the dynamics at the small scale, 
which affects the result of the spectrum of gravitational waves.
As a check, we performed test simulations by using the modified equation of motion suggested by~\cite{Press:1989yh,Sousa:2010zza},
and confirmed that the result for the spectrum of gravitational waves becomes different compared with that obtaind
from the simulation with the canonical equation of motion~\eqref{eq3-1},
even though the result for the scaling parameter~\eqref{eq2-4} is the same in each scheme.

\subsection{\label{sec3-1}Initial conditions}
The initial field configuration is generated in a similar way to the previous study~\cite{Kawasaki:2011vv}.
In the momentum space, we give the initial value of the scalar field and its time derivative
as Gaussian random fields, such that they satisfy the correlation function induced by quantum fluctuation of the massless scalar field:
\begin{align}
\langle\phi({\bf k})\phi({\bf k'})\rangle &= \frac{1}{2k}\delta^{(3)}({\bf k+k'}), \nonumber\\
\langle\dot{\phi}({\bf k})\dot{\phi}({\bf k'})\rangle &= \frac{k}{2}\delta^{(3)}({\bf k+k'}). \label{eq3-2}
\end{align}
Since the spectrum of the field diverges at large $k$, we truncate the modes whose wavenumber is higher than a cutoff scale $k_{\rm cut}$,
and treat $k_{\rm cut}$ as the input parameter of the numerical simulation.
The large initial field fluctuations at large $k$ would affect the result of the numerical simulation since the simulation time is
not so long enough for the initial fluctuations to be diluted.
A simple way to avoid the contamination from the initial field fluctuations is to use the Gaussian initial condition~\eqref{eq3-2}
with a sufficiently small value of $k_{\rm cut}$, which suppresses the unexpected feature on the gravitational wave spectrum at large $k$.
We compared the result of the numerical simulation with varying the value of $k_{\rm cut}$
and confirmed that the dependence on the choice of the initial conditions hardly appears 
as long as we use the value $k_{\rm cut}\lesssim 1$.
Regarding this fact, we use this Gaussian initial condition with $k_{\rm cut}=1$ in the results shown in this paper.

\subsection{\label{sec3-2}Scaling parameter}
In each realization of the simulation,
we checked the scaling property represented by Eq.~\eqref{eq2-3}.
This property can be investigated in two ways:
One is to compute the area of the surface of domain walls in the simulation box and check the constancy of the scaling parameter
$\mathcal{A}$ defined in Eq.~\eqref{eq2-4}.
The other is to follow the evolution of the energy density of domain walls $\rho_{\rm wall}$, which can be calculated from the data of the scalar field in the simulation box.
In this work we carry out both ways to confirm the scaling property of domain walls.

In numerical simulations, the scaling parameter $\mathcal{A}$ is computed as follows.
Let us denote the area of domain walls in the comoving coordinate as $A$, and the volume of the comoving simulation box as $V$.
Since the energy of domain walls existing within the simulation box becomes $\sigma_{\rm wall}a^2(t)A$,
the energy density is given by
\begin{equation}
\rho_{\rm wall} = \frac{\sigma_{\rm wall}A}{a(t)V}. \label{eq3-3}
\end{equation}
Substituting it into Eq.~\eqref{eq2-4}, we obtain
\begin{equation}
\mathcal{A} = \frac{At}{a(t)V}. \label{eq3-4}
\end{equation}

The area of domain walls $A$ is computed by the use of the algorithm introduced in Ref.~\cite{Press:1989yh}.
We compare the sign of the scalar field $\phi$ between two neighboring grid points (let us call it as a link),
defining the quantity $\delta_{\pm}$ which takes $1$ if the sign changes on the link and $0$ otherwise.
Then, we sum up $\delta_{\pm}$ over all links in the simulation box with multiplying a weighting factor:
\begin{equation}
A = \Delta A \sum_{\rm links} \delta_{\pm}\frac{|\nabla\phi|}{|\phi_{,x}|+|\phi_{,y}|+|\phi_{,z}|}, \label{eq3-5}
\end{equation}
where $\Delta A = (\Delta x)^2$ is the area of one grid surface, $\Delta x$ is the spacing between two neighboring
grid points, and $\phi_{,i}\ (i=x,y,z)$ are the spatial derivatives of $\phi({\bf x})$.
The weighting factor takes account of the average number of links per area segment. 
This calculation scheme is used in the other literature to estimate the scaling property of domain walls~\cite{Press:1989yh,Oliveira:2004he}.

To check the scaling law, we also estimate $\rho_{\rm wall}$ directly without using Eq.~\eqref{eq3-3}.
Here, the average of the energy density of domain walls in the simulation box is evaluated from the total energy density
of the scalar field inside the core of them:
\begin{equation}
\rho_{\rm wall}(t) = \frac{1}{N^3}\sum_{\rm grids}\rho_{\rm wall}({\bf x},t), \label{eq3-6}
\end{equation}
where a term ``grids" indicates the summation over all grid points, and $\rho_{\rm wall}({\bf x},t)$ is given by
\begin{equation}
\rho_{\rm wall}({\bf x},t) = \left\{
\begin{array}{l l}
\frac{1}{2}\dot{\phi}^2 + \frac{1}{2a^2}|\nabla\phi|^2 + V(\phi) & \mathrm{inside\ the\ core\ of\ walls}, \\
0 & \mathrm{otherwise}.
\end{array}\right. \label{eq3-7}
\end{equation}
The region of the core of walls is identified as follows.
Firstly, we identify the grid points on which the link with $\delta_{\pm}=1$ is attached, and call them as the ``loci of walls".
Then, we increment the grid points within the distance $d=2\delta_w=(2\sqrt{2})/(\sqrt{\lambda}\eta)$ from the loci of walls to evaluate the right-hand-side of Eq.~\eqref{eq3-7},
where $\delta_w\simeq (\sqrt{\lambda/2}\eta)^{-1}$ is the width of domain walls.

\subsection{\label{sec3-3}Spectrum of gravitational waves}
The spectrum of gravitational waves is computed by using the method developed by~\cite{Dufaux:2007pt},
which was originally introduced to estimate the gravitational waves produced at preheating after inflation~\cite{Khlebnikov:1997di,Dufaux:2007pt}.
This method enables us to compute the spectrum of gravitational waves at a given time $t$ as a time integral of the source term
constructed from the spatial derivative of the scalar field convoluted with the homogeneous solutions of the equations for metric perturbations.
 
Here we use the notations of Ref.~\cite{Hiramatsu:2012sc} 
and briefly summarize formulae for the power spectrum of gravitational waves.
The spectrum of gravitational waves at the time $t$ is given by
\begin{align}
\frac{d\rho_{\rm gw}}{d\ln k}(t) &= \frac{G}{2\pi^2 Va^4(t)}S_k(t), \label{eq3-8} \\
S_k(t) &= k\int d\Omega_k \sum_{ij}\left(\left|\bar{C}_{ij}^{(1)}\right|^2 + \left|\bar{C}_{ij}^{(2)}\right|^2 \right), \label{eq3-9}
\end{align}
where $d\Omega_k=d\cos\theta d\phi$ represents the measure of the integration over the direction of ${\bf k}$,
and $\bar{C}^{(1)}_{ij}$ and $\bar{C}^{(1)}_{ij}$ are given by
\begin{align}
\bar{C}^{(1)}_{ij}({\bf k},\tau) &= -\frac{k}{16\pi G}\int^{\tau}_{\tau_i}d\tau'\sin k\tau'S_{ij}({\bf k},\tau'), \nonumber\\
\bar{C}^{(2)}_{ij}({\bf k},\tau) &= \frac{k}{16\pi G}\int^{\tau}_{\tau_i}d\tau'\cos k\tau'S_{ij}({\bf k},\tau'). \label{eq3-10}
\end{align}
In Eq.~\eqref{eq3-10}, we introduced the conformal time $\tau$ defined by $d\tau=dt/a$, and $\tau_i$ represents the initial time
of the numerical simulation.
$S_{ij}$ is the source function defend as follows
\begin{align}
S_{ij}({\bf k},\tau) = 16\pi Ga(\tau)T_{ij}^{\rm TT}({\bf k},\tau). \label{eq3-11}
\end{align}
$T^{\rm TT}_{ij}$ is constructed from the transverse traceless (TT) projection of the stress-energy tensor $T_{ij}$ of the scalar field:
\begin{align}
T_{ij}^{\rm TT}({\bf k},\tau) &= \Lambda_{ij,kl}(\hat{k})T_{kl}({\bf k},\tau) = \Lambda_{ij,kl}\{\partial_k\phi\partial_l\phi\}({\bf k},\tau), \label{eq3-12} \\
\Lambda_{ij,kl}(\hat{k}) &= P_{ik}(\hat{k})P_{jl}(\hat{k}) - \frac{1}{2}P_{ij}(\hat{k})P_{kl}(\hat{k}), \label{eq3-13} \\
P_{ij}(\hat{k}) &= \delta_{ij} - \hat{k}_i\hat{k}_j, \label{eq3-14} \\
\hat{k} &= {\bf k}/|{\bf k}|, \label{eq3-15}
\end{align}
where $\{\partial_k\phi\partial_l\phi\}({\bf k},\tau)$ represents the Fourier transform of $\partial_k\phi({\bf x},\tau)\partial_l\phi({\bf x},\tau)$.

In addition to the estimation of the spectrum of gravitational waves,
we also compute the efficiency parameter defined in Eq.~\eqref{eq2-6}.
Integrating Eq.~\eqref{eq3-8}, we obtain the total energy density of gravitational waves:
\begin{equation}
\rho_{\rm gw}(t) = \frac{G}{2\pi^2 V a^4(t)}\int\frac{dk}{k}S_k(t). \label{eq3-16}
\end{equation}
From Eqs.~\eqref{eq2-6} and~\eqref{eq3-16}, the efficiency parameter can be estimated as
\begin{equation}
\epsilon_{\rm gw} = \frac{1}{2\pi^2 V\mathcal{A}^2\sigma_{\rm wall}^2 a^4(t)}\int \frac{dk}{k}S_k(t). \label{eq3-17}
\end{equation}

It should be noted that the normalization of the spectrum of gravitational waves $d\rho_{\rm gw}/d\ln k$
can be fixed such that it reproduces the value of $\epsilon_{\rm gw}$ obtained from numerical simulations
when we take the integral of the quantity $(G\mathcal{A}^2\sigma_{\rm wall}^2)^{-1}(d\ln\rho_{\rm gw}/d\ln k$) over $\ln k$.
However, as we will see in Sec.~\ref{sec4-2}, the integration procedure contains a large systematic uncertainty, and
it would be practically difficult to fix the normalization of the spectrum by using the integrated quantity such as $\epsilon_{\rm gw}$.
Therefore, instead of using $\epsilon_{\rm gw}$ to determine the normalization of the spectrum of gravitational waves,
we will use the differential amplitude defined as follows
\begin{equation}
\tilde{\epsilon}_{\rm gw}\equiv \left(\frac{d\epsilon_{\rm gw}}{d\ln k}\right)_{\rm peak} = \frac{\left.S_k(t)\right|_{\rm peak}}{2\pi^2V\mathcal{A}^2\sigma_{\rm wall}^2a^4(t)}, \label{eq3-18}
\end{equation}
where the subscript ``peak" implies that the value is computed at the peak of the spectrum $S_k$.
Since this quantity is defined at the peak wavenumber, we expect that the value of $\tilde{\epsilon}_{\rm gw}$
also approaches a constant in the scaling regime.
We will check the constancy of $\tilde{\epsilon}_{\rm gw}$ and estimate its value in Sec.~\ref{sec4-2}.

\subsection{\label{sec3-4}Parameters used in the simulations}
Let us make a comment on the parameters used in the numerical simulations.
In numerical studies we normalize all dimensionful quantities in the unit of $\eta=1$.
In this unit, denoting the grid size and the box size as $N^3$ and $V=L^3$, respectively,
we performed simulations for two cases, $(N,L)=(512,80)$ and $(1024,120)$.
In both cases the time integration is performed with the interval $d\tau=0.02$ in the conformal time.
The time evolution of the scale factor is determined such that it follows the expansion in the 
radiation dominated universe ($a\propto t^{1/2}\propto \tau$).
The normalization of the scale factor is taken such that $a(\tau_i)=1$ at the initial time of the simulation,
and we fix the initial time as $\tau_i=2$.
The final time of the simulation is chosen as $\tau_f=40$ for simulations with $N=512$
and $\tau_f=60$ for those with $N=1024$.
Hence the dynamical range of the simulation becomes $\tau_f/\tau_i=20$ for $N=512$
and $\tau_f/\tau_i=30$ for $N=1024$.
Note that the dynamical ranges of the current simulations are much larger than
that of the previous studies~\cite{Hiramatsu:2010yz,Kawasaki:2011vv}, where we used $N=256$ and $\tau_f/\tau_i\simeq 12$.

The other theoretical parameter is the coupling constant $\lambda$ in the potential~\eqref{eq2-2},
which controls the surface mass density $\sigma_{\rm wall} = 2\sqrt{2\lambda}\eta^3/3$ and the width $\delta_w \simeq (\sqrt{\lambda/2}\eta)^{-1}$ of domain walls.
Note that the value of this parameter is constrained from the requirements that the Hubble radius must be shorter than the box size,
and that the width of domain walls must be larger than the physical lattice spacing $\Delta x_{\rm phys} = a(t)L/N$.
The scale of the Hubble radius and the width of domain walls divided by the physical lattice spacing are given by
\begin{equation}
\frac{H^{-1}}{\Delta x_{\rm phys}} = \frac{N}{L}\tau \qquad {\rm and} \qquad \frac{\delta_w}{\Delta x_{\rm phys}} = \frac{N}{L\sqrt{\lambda/2}}\left(\frac{\tau_i}{\tau}\right). \label{eq3-19}
\end{equation}
We require that at least the conditions $H^{-1}/\Delta x_{\rm phys}<N$ and $\delta_w/\Delta x_{\rm phys}>1$ must be satisfied at the end of the simulation $\tau=\tau_f$.
In the previous studies~\cite{Hiramatsu:2010yz,Kawasaki:2011vv}, we fixed
the value of $\lambda$ with satisfying the above conditions
due to the limitation of the dynamical range of the simulation.
On the other hand, in the current simulations we can vary the value of $\lambda$ and compare the results
with different values of $\lambda$.
Here, we perform simulations for three choices of the parameter, $\lambda=0.03$, $0.01$, and $0.003$.

The parameters used in the simulations are summarized in Table~\ref{tab1}.
For each value of $\lambda$, we execute 10 realizations with $N=512$ and 1 realization with $N=1024$.
Before calculating the quantities such as the scaling parameter and the spectrum of gravitational waves,
we take an average over 10 realizations for the simulations with $N=512$.
Hence we perform simulations in 6 different sets of parameters as shown in Table~\ref{tab1} [case (a) to (f)].
Note that the requirements described below Eq.~\eqref{eq3-19} are satisfied for every set of parameters.
We also note that the dependence on the bias parameter $\epsilon$ appearing in Eq.~\eqref{eq2-7} is not investigated
in the numerical study, since we do not simulate the collapse of domain walls.
Effects of the different choice of parameters on the numerical results will be discussed in the next section.

\begin{table}[h]
\begin{center} 
\caption{Different sets of parameters used in numerical simulations.}
\vspace{3mm}
\begin{tabular}{c c c c c c c}
\hline\hline
Case & Grid size ($N^3$) & Box size ($L$) & $\lambda$ & $H^{-1}/\Delta x_{\rm phys}$ (at $\tau=\tau_f$) & $\delta_w/\Delta x_{\rm phys}$ (at $\tau=\tau_f$) & Realization \\
\hline 
(a) & $512^3$ & 80 & 0.03 & 256 & 2.61 & 10 \\
(b) & $512^3$ & 80 & 0.01 & 256 & 4.53 & 10 \\
(c) & $512^3$ & 80 & 0.003 & 256 & 8.26 & 10 \\
(d) & $1024^3$ & 120 & 0.03 & 512 & 2.32 & 1 \\
(e) & $1024^3$ & 120 & 0.01 & 512 & 4.02 & 1 \\
(f) & $1024^3$ & 120 & 0.003 & 512 & 7.34 & 1 \\
\hline\hline
\label{tab1}
\end{tabular}
\end{center}
\end{table}

\section{\label{sec4}Results}
\subsection{\label{sec4-1}Scaling properties}

Figure 1 shows the evolution of the area parameter $\mathcal{A}$ defined in Eq.~\eqref{eq2-4}.
We see that the value of $\mathcal{A}$ becomes almost constant at late times, 
which implies that the network of domain walls
obeys the scaling law at the late time of the simulations.
We note that the value of $\mathcal{A}$ does not depend on the choice of $\lambda$.
Approaching to the constant value is observed more obviously in the simulations with $N=1024$ than those with $N=512$,
except for the fact that the value of $\mathcal{A}$ slightly increases around the final time of the simulations.
It is presumed that this deviation from the constancy is caused by the finiteness of the simulation box, since the Hubble radius becomes
comparable to the box size at the final time of the simulations.
This effect would be removed if we performed the simulation with a larger box size.

As shown in the plot for the simulations with $N=512$, there is a statistical uncertainty of $\mathcal{O}(0.1)$
due to the different realizations. Including this uncertainty, we estimate the value of $\mathcal{A}$
in the scaling regime as
\begin{equation}
\mathcal{A} \simeq 0.8\pm0.1. \label{eq4-1}
\end{equation}

\begin{figure}[htbp]
\begin{center}
\includegraphics[scale=1.0]{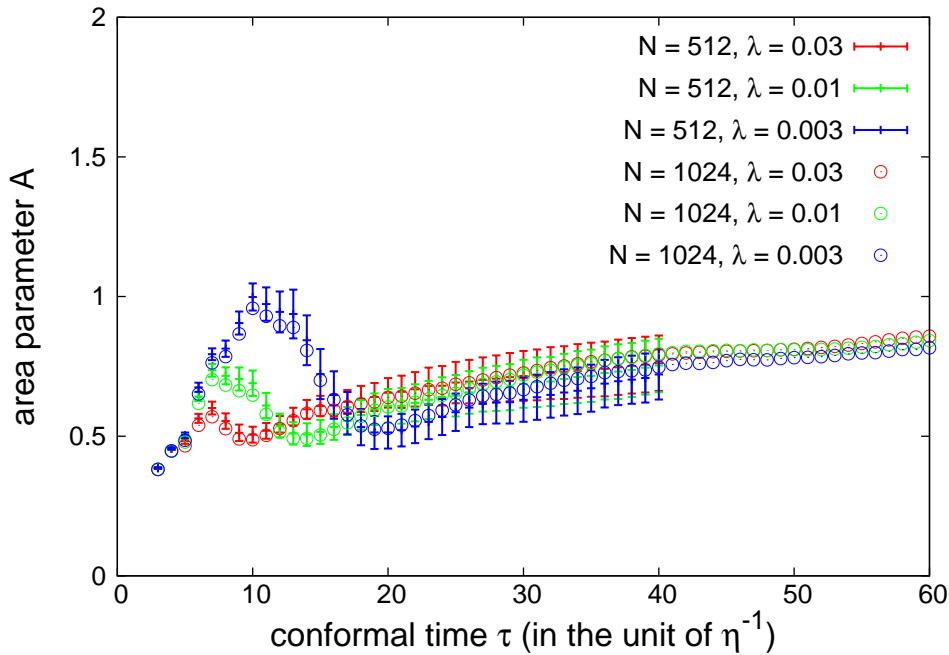}
\end{center}
\caption{Time evolution of the area parameter given by Eq.~\eqref{eq3-4}. Here, we plot the results of the simulations
with the sets of parameters shown in (a)-(f) of Table~\ref{tab1}.}
\label{fig1}
\end{figure}

\begin{figure}[htbp]
\begin{center}
\includegraphics[scale=1.0]{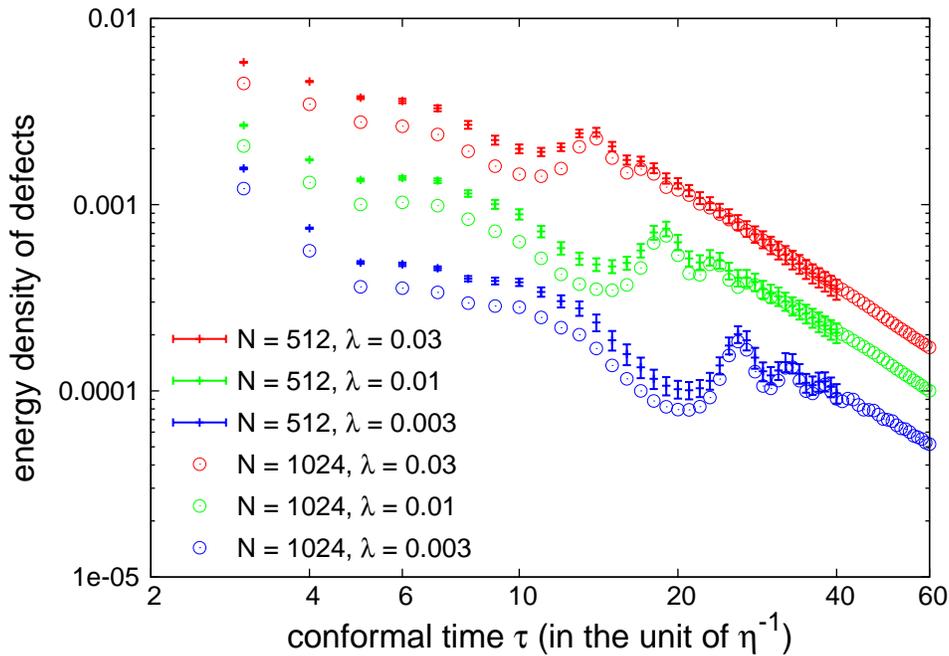}
\end{center}
\caption{Time evolution of the energy density of domain walls for the sets of parameters shown in (a)-(f) of Table~\ref{tab1}.
The energy density is computed from Eqs.~\eqref{eq3-6} and~\eqref{eq3-7} with a criterion to determine the core region of the walls
described in the text below Eq.~\eqref{eq3-7}.}
\label{fig2}
\end{figure}

We also plot the time evolution of the energy density of domain walls in Fig.~\ref{fig2}.
We confirm that $\rho_{\rm wall}$ evolves along to the scaling law $\propto t^{-1} \propto \tau^{-2}$ at late times.
The magnitude of the energy density scales as $\sqrt{\lambda}$, which coincides with the relations
$\rho_{\rm wall} = \mathcal{A}\sigma_{\rm wall}/t$ and $\sigma_{\rm wall}\propto \sqrt{\lambda}$.
We also observe that the onset of scaling (or the formation time $\tau_{\rm form}$ of the horizon-sized walls) depends on $\lambda$.
This formation time scale might be estimated from the condition that
the width of walls $\delta_w\simeq (\sqrt{\lambda/2}\eta)^{-1}$ becomes comparable (or shorter than) the Hubble radius $H^{-1}$.
This condition gives $\tau_{\rm form}\propto \lambda^{-1/4}$, as shown in Fig.~\ref{fig2}.

We note that the disagreement between the results for $N=512$ and those for $N=1024$ at the early times of the simulations in Fig.~\ref{fig2}
is caused by the fact that we cannot correctly determine the core region of domain walls before they enter into the scaling regime.
At the initial time, the scalar field distributes according to the initial condition described in Sec.~\ref{sec3-1},
and it varies with an interval $\sim k_{\rm cut}^{-1}$. Since we identify the loci of walls as the place where
the scalar field changes its sign, as long as the lattice spacing $\Delta x$ is shorter than $k_{\rm cut}^{-1}$,
$\sum_{\rm grids}\rho_{\rm wall}({\bf x},t)$ in Eq.~\eqref{eq3-6} does not change so much for fixed value of $k_{\rm cut}$
at the initial time (though it depends on the simulation box size), while $\rho_{\rm wall}(t)$ in the left-hand side of Eq.~\eqref{eq3-6}
is suppressed by the factor $N^{-3}$. This is the reason why the result of $\rho_{\rm wall}$ for $N=1024$ is
smaller than that for $N=512$ at the initial stage of the simulation.
It should be emphasized that this difference at the initial stage due to the numerical artifacts in the initial conditions is a matter of minor importance
for our purpose, since we are interested in the evolution of domain walls during the scaling regime.
Indeed, from Fig.~\ref{fig2} we see that the difference of the results between simulations with $N=512$ and $N=1024$
is not prominent after walls enter into the scaling regime.

\subsection{\label{sec4-2}Gravitational waves}

\begin{figure}[htbp]
\begin{center}
\includegraphics[scale=1.0]{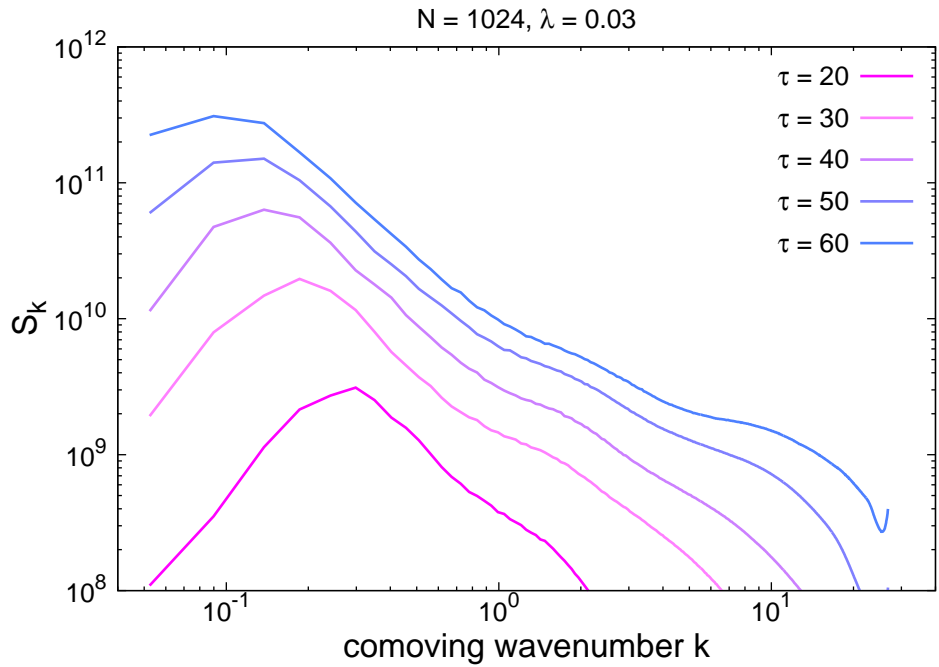}
\end{center}
\caption{Time evolution of the function $S_k$ defined in Eq.~\eqref{eq3-9} for the simulation with $N=1024$ and $\lambda=0.03$
[case (d) in Table.~\ref{tab1}].
The spectra are plotted from $\tau=20$ (pink) to $\tau=60$ (blue) in the conformal time with the interval $\Delta\tau=10$.}
\label{fig3}
\end{figure}

Next, let us show the results for the spectrum of gravitational waves.
In Fig.~\ref{fig3}, we plot the time evolution of the spectrum $S_k$ obtained form the simulation with $N=1024$ and $\lambda=0.03$.
This result indicates that the spectrum peaks at small $k$ while the amplitude is suppressed at large $k$.
The form of $S_k$ shown in Fig.~\ref{fig3} differs from the previous result~\cite{Hiramatsu:2010yz,Kawasaki:2011vv}
in which the spectrum becomes rather flat in the intermediate scale.
As we describe in Appendix~\ref{secA}, this discrepancy is due to the fatal error in numerical codes used in the previous studies.

\begin{figure}[htbp]
\begin{center}
\includegraphics[scale=1.0]{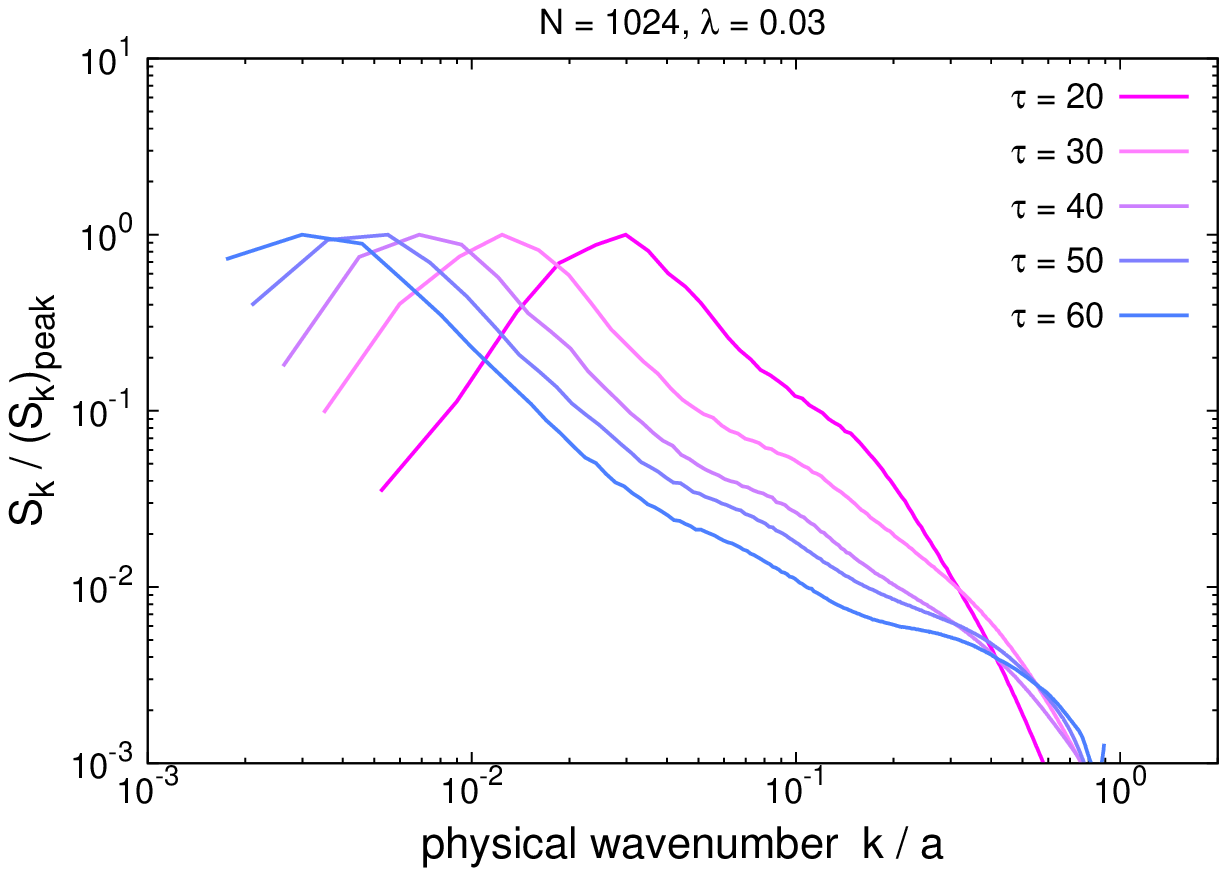}
\end{center}
\caption{Time evolution of the function $S_k$ for the simulation with $N=1024$ and $\lambda=0.03$
[case (d) in Table.~\ref{tab1}] with taking horizontal axis as the physical wavenumber $k/a$ and normalizing
vertical axis by the peak amplitude. 
The spectra are plotted from $\tau=20$ (pink) to $\tau=60$ (blue) in the conformal time with the interval $\Delta\tau=10$.}
\label{fig4}
\end{figure}

\begin{figure*}[htp]
\centering
$\begin{array}{cc}
\subfigure[]{
\includegraphics[width=0.45\textwidth]{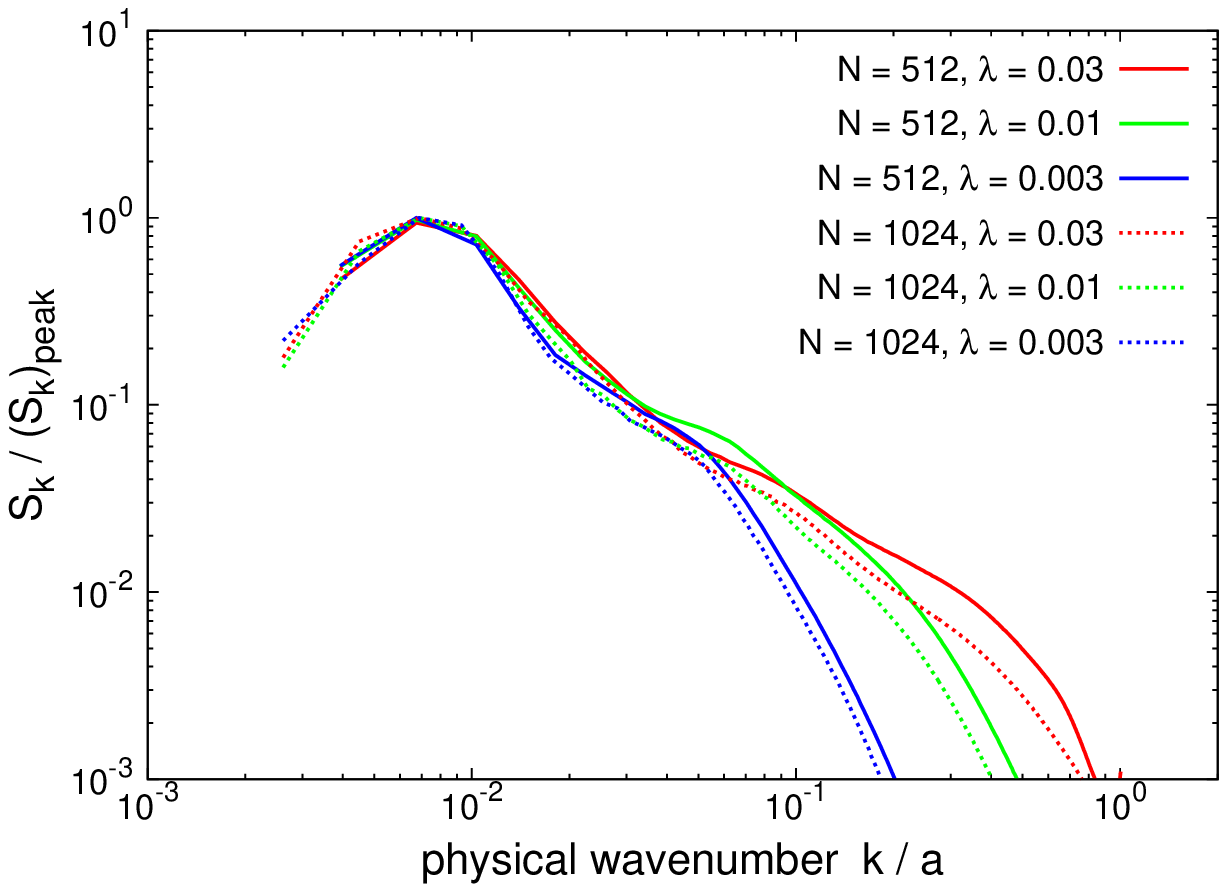}}
\hspace{20pt}
\subfigure[]{
\includegraphics[width=0.45\textwidth]{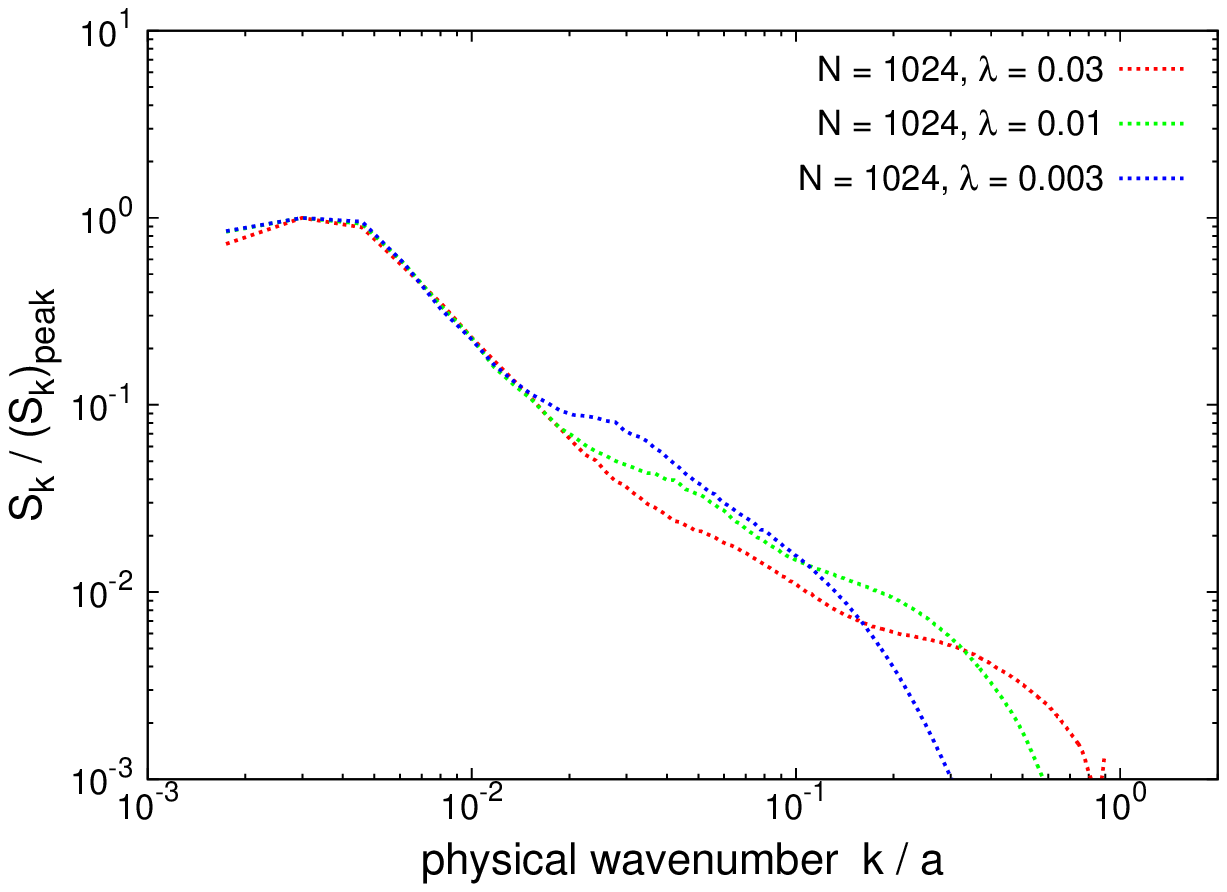}}
\end{array}$
\caption{The Spectra $S_k$ normalized by the peak amplitude for various values of $\lambda$
with taking horizontal axis as the physical wavenumber $k/a$.
Left panel (a) shows the results at the conformal time $\tau=40$ for the sets of parameters (a)-(f) in Table~\ref{tab1},
and right panel (b) shows those at $\tau=60$ for the sets of parameters (d)-(f) in Table~\ref{tab1}.
Here, the results of the simulations with $N=512$ are shown as solid lines, and those with $N=1024$ are shown as broken lines.
}
\label{fig5}
\end{figure*}

We also plot the evolution of the spectrum with taking the horizontal axis as the physical wavenumber in Fig.~\ref{fig4}.
This figure indicates that the peak is located at the horizon scale, which shifts with time as $k/a\propto H$.
On the other hand, we also observe that the spectrum always falls off at a large wavenumber $k/a \sim 0.5$,
whose location does not seem to depend on time.
We deduce that this falloff corresponds to the microscopic scale determined by
the width of domain walls ($k/a\sim \delta_w^{-1}\simeq \sqrt{\lambda/2}\eta$).
This speculation can be checked in part when we compare the results for different values of $\lambda$, as shown in Fig.~\ref{fig5}.
We see that the falloff of the spectrum occurs at higher momentum region for the result with larger value of $\lambda$.
This result is consistent with the above speculation that the spectrum extends up to the scale $\delta_w^{-1}\propto \sqrt{\lambda}$.
On the other hand, the peak location scarcely depends on the value of $\lambda$, from which we confirm that the peak is determined
by the Hubble parameter. The slope of the spectrum in the intermediate scale between the Hubble radius and the width of walls
is not determined straightforwardly, but from the results with the larger dynamical range [Fig.~\ref{fig5} (b)] we observe that $S_k$
decreases as $k^{-1}$ at large wavenumbers.

\begin{figure*}[htp]
\centering
$\begin{array}{cc}
\subfigure[]{
\includegraphics[width=0.45\textwidth]{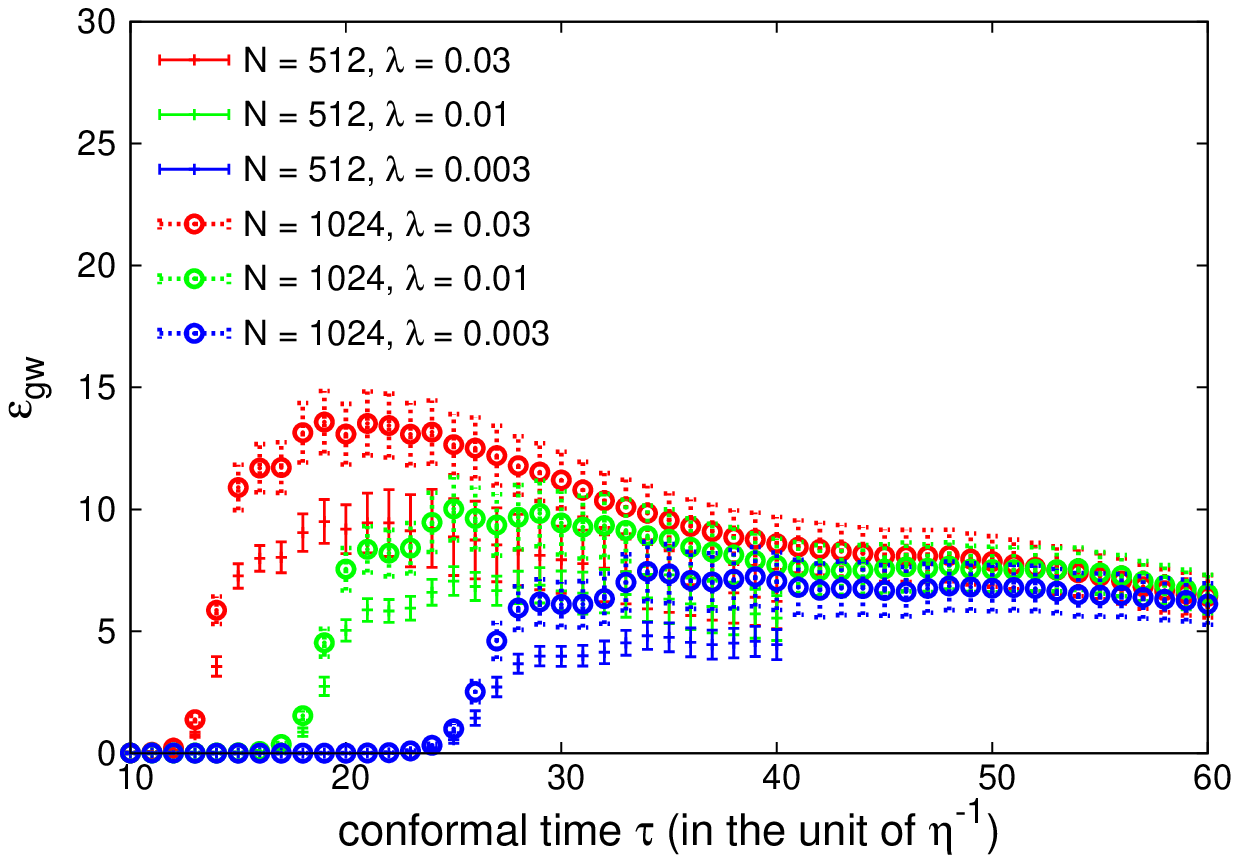}}
\hspace{20pt}
\subfigure[]{
\includegraphics[width=0.45\textwidth]{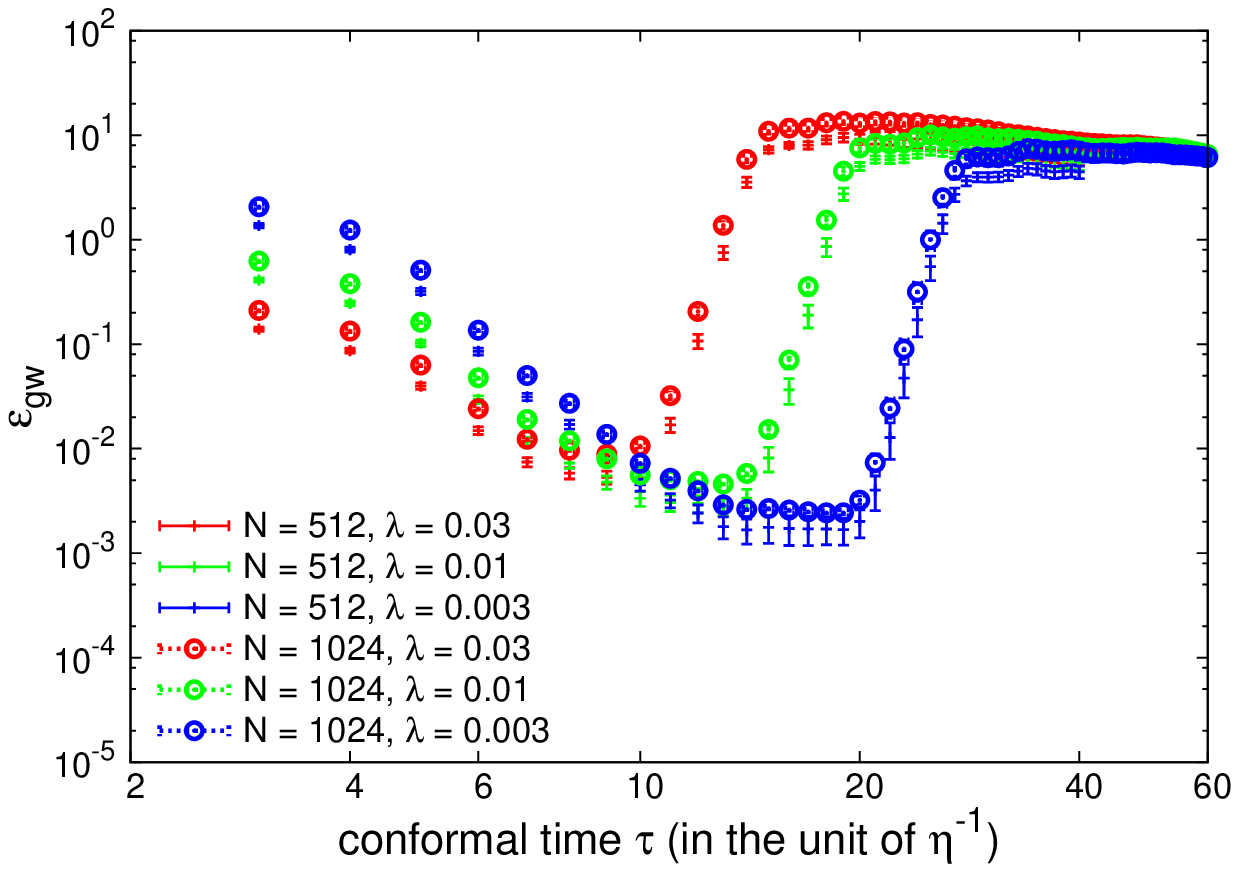}}\\
\subfigure[]{
\includegraphics[width=0.45\textwidth]{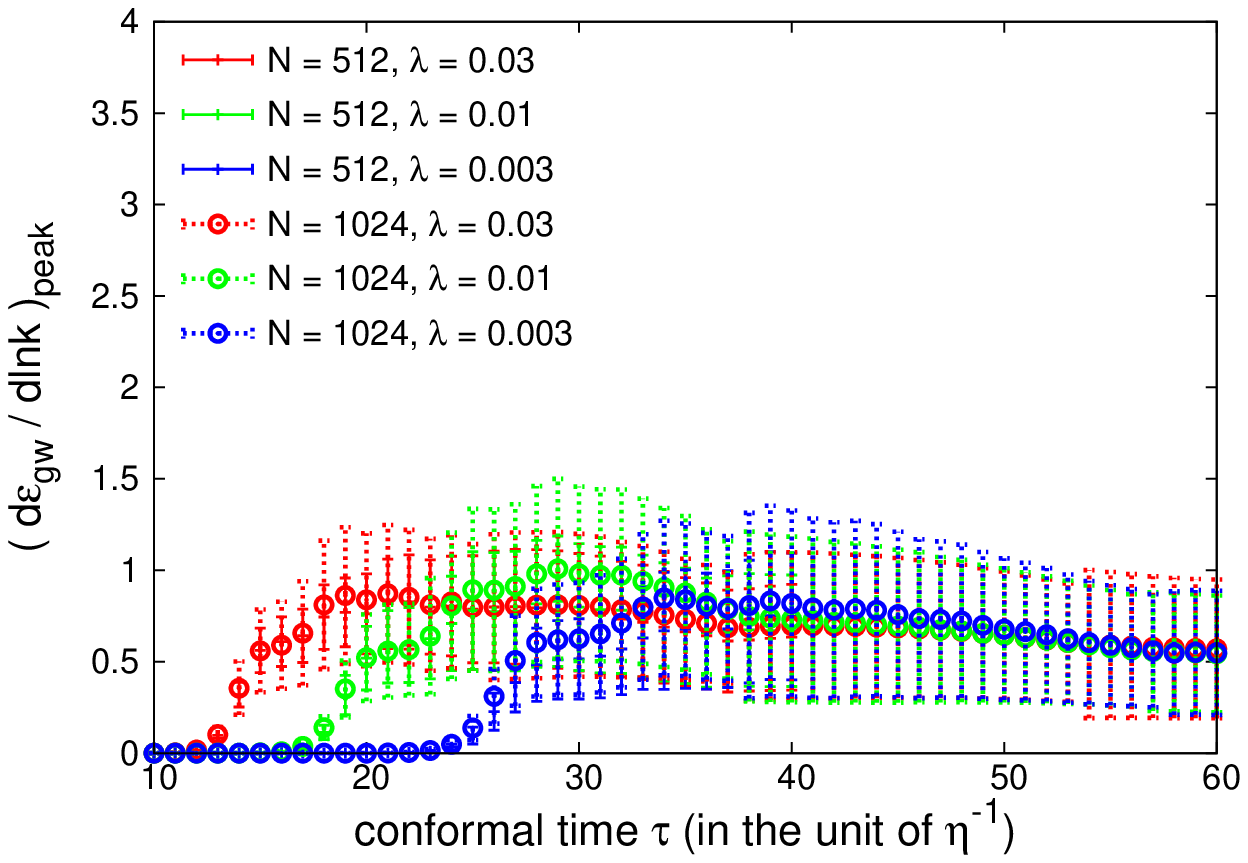}}
\hspace{20pt}
\subfigure[]{
\includegraphics[width=0.45\textwidth]{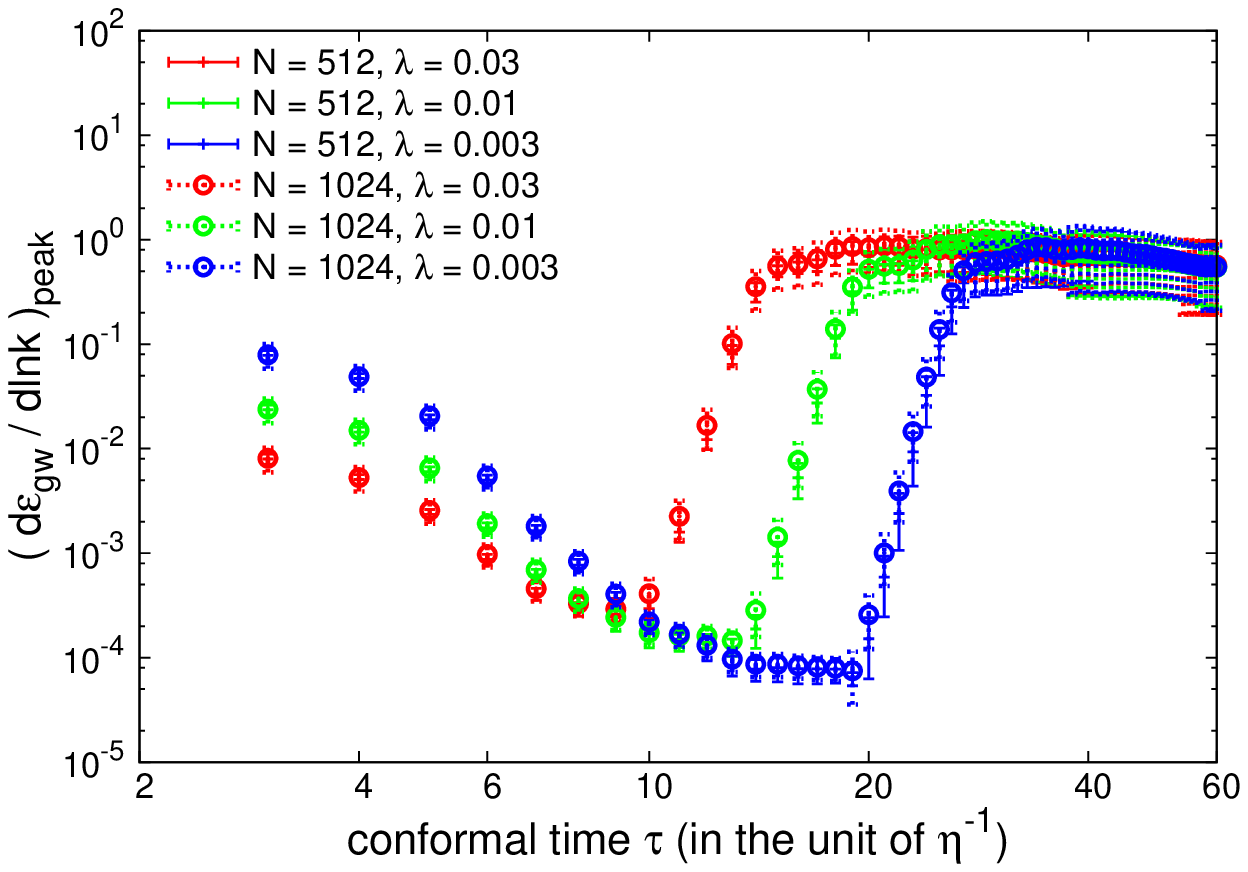}}
\end{array}$
\caption{Time evolution of the efficiency parameter $\epsilon_{\rm gw}$ given by Eq.~\eqref{eq3-17} and its differentiation at the peak
$(\epsilon_{\rm gw}/\ln k)_{\rm peak}$ given by Eq.~\eqref{eq3-18} for the simulations
with the sets of parameters shown in (a)-(f) of Table~\ref{tab1}.
Here, we plot $\epsilon_{\rm gw}$ in (a) linear scale and (b) logarithmic scale, and also $(\epsilon_{\rm gw}/\ln k)_{\rm peak}$
in (c) linear scale and (d) logarithmic scale.}
\label{fig6}
\end{figure*}

Figures~\ref{fig6} (a) and (b) show the evolution of the efficiency parameter defined in Eq.~\eqref{eq3-17}.
From these figures we see that the magnitude of $\epsilon_{\rm gw}$ rapidly grows at the time of the formation of domain walls,
and it remains almost constant value of $\mathcal{O}(1)$ at late times.
It should be noted that there are various uncertainties on the determination of the value of $\epsilon_{\rm gw}$.
First of all, there are two statistical errors shown as error bars in figures. One is the variation for each realization of the simulation.
This statistical error is not included in the results of the simulations with $N=1024$, since we performed one realization only.
The other kind of error arises when we evaluate the integration over the direction of ${\bf k}$ to obtain the spectrum $S_k$.
In numerical analysis, we compute $\int d\Omega_k$ in Eq.~\eqref{eq3-9} by averaging all data defined on the shell with $|{\bf k}|=k$
in $k$-space. We estimate the statistical error as a standard deviation arising in this averaging procedure.
In addition to these statistical errors, there is a systematic error,
which appears in the difference in the results between the simulation with $N=512$ and that with $N=1024$ [see Figs.~\ref{fig6} (a) and (b)].
This difference occurs due to the poor resolution of the peak of the spectrum (i.e. the small number of bins at small $k$)
when we integrate $S_k/k$ over $k$ to obtain $\epsilon_{\rm gw}$ by using Eq.~\eqref{eq3-17}.
Since there is a few data points in small $k$ and the peak of $S_k$ is located there,
we are apt to underestimate the total amplitude if the grid number $N$ is small.

To check that the systematic error described above arises due to the integration over $k$ with discretized grids,
we also compute the
differential amplitude $\tilde{\epsilon}_{\rm gw}= (d\epsilon_{\rm gw}/d\ln k)_{\rm peak}$ given by Eq.~\eqref{eq3-18}.
Since this quantity is just determined by the peak amplitude, we expect that its value is unaffected by the uncertainty in the discretization scheme.
Figures~\ref{fig6} (c) and (d) show the evolution of $(d\epsilon_{\rm gw}/d\ln k)_{\rm peak}$, from which we confirm that the results do not depend on the choice of $N$,
and that it takes almost constant value in the scaling regime.
However, we also see that the statistical error becomes large ($\sim 50\mathchar`-60\%$) in these plots.

As was noted in Sec.~\ref{sec3-3}, we can fix the normalization of the spectrum by using the value $\tilde{\epsilon}_{\rm gw}$ determined from numerical simulations.
From the average of six data points plotted at $\tau=40$ in Fig.~\ref{fig6} (c), we obtain
\begin{equation}
\tilde{\epsilon}_{\rm gw} \simeq 0.7\pm 0.4, \label{eq4-2}
\end{equation}
where we took a conservative error ($60$\% of the mean value), regarding the large statistical uncertainties.
We will use this value to estimate the present density of gravitational waves in Sec.~\ref{sec5}.

\section{\label{sec5} Present density of gravitational waves}
In this section, we estimate the amplitude and frequency of gravitational waves observed today
by using the numerical results described in the previous section.
The amplitude can be estimated from the fact that the energy density of gravitational waves remains almost constant during the scaling regime
\begin{equation}
\rho_{\rm gw} = \epsilon_{\rm gw}G\mathcal{A}^2\sigma_{\rm wall}^2. \label{eq5-1}
\end{equation}
Alternatively, we can estimate the amplitude at the peak of the spectrum in terms of the quantity $\tilde{\epsilon}_{\rm gw}$ defined in Eq.~\eqref{eq3-18}:
\begin{equation}
\left(\frac{d\rho_{\rm gw}}{d\ln k}\right)_{\rm peak} = \tilde{\epsilon}_{\rm gw}G\mathcal{A}^2\sigma_{\rm wall}^2. \label{eq5-2}
\end{equation}
Assuming that the gravitational radiation terminates at the decay time of domain walls ($t=t_{\rm dec}$),
and that the decay completes in the radiation dominated era,
we estimate the fraction between the energy density of gravitational waves and the total energy density of the universe $\rho_c(t_{\rm dec})=3H^2(t_{\rm dec})/8\pi G$
at the decay time as
\begin{equation}
\Omega_{\rm gw}(t_{\rm dec})_{\rm peak} \equiv \frac{1}{\rho_c(t_{\rm dec})}\left(\frac{d\rho_{\rm gw}}{d\ln k}\right)_{\rm peak} = \frac{8\pi\tilde{\epsilon}_{\rm gw}G^2\mathcal{A}^2\sigma_{\rm wall}^2}{3H^2(t_{\rm dec})}, \label{eq5-3}
\end{equation}
where 
\begin{equation}
H(t_{\rm dec}) = \frac{1}{2t_{\rm dec}} \simeq \sqrt{\frac{2}{\lambda}}\epsilon\eta \label{eq5-4}
\end{equation}
is the Hubble parameter at $t=t_{\rm dec}$.
If the production of gravitational waves completes in the radiation dominated era, the present density of gravitational wavs is estimated as
\begin{equation}
\Omega_{\rm gw}h^2(t_0) = \Omega_Rh^2\left(\frac{g_0}{g_*}\right)^{1/3}\Omega_{\rm gw}(t_{\rm dec}) = 1.34\times 10^{-5}\left(\frac{100}{g_*}\right)^{1/3}\Omega_{\rm gw}(t_{\rm dec}), \label{eq5-5}
\end{equation}
where $t_0$ represents the present time, 
$h$ is the reduced Hubble parameter ($H_0=100h\ {\rm km\cdot sec^{-1}Mpc^{-1}}$), 
$\Omega_Rh^2=4.15\times 10^{-5}$ is the density parameter of radiations at the present time, and
$g_0= 3.36$ and $g_*$ are the relativistic degrees of freedom at the present time and at
$t=t_{\rm dec}$, respectively.
Combining Eqs.~\eqref{eq5-3}-\eqref{eq5-5}, we obtain
\begin{align}
\Omega_{\rm gw}h^2(t_0)_{\rm peak} &\simeq 2.2\times 10^{-21}\times \tilde{\epsilon}_{\rm gw}\mathcal{A}^2\lambda^2\epsilon^{-2}\left(\frac{g_*}{100}\right)^{-1/3}
\left(\frac{\eta}{10^{15}\mathrm{GeV}}\right)^4 \label{eq5-6}\\
&\simeq 1.0 \times 10^{-21}\times\lambda^2\epsilon^{-2}\left(\frac{g_*}{100}\right)^{-1/3}\left(\frac{\eta}{10^{15}\mathrm{GeV}}\right)^4, \label{eq5-7}
\end{align}
where we used the mean values of $\mathcal{A}$ and $\tilde{\epsilon}_{\rm gw}$ obtained from
the numerical results, $\mathcal{A}\simeq 0.8$ and $\tilde{\epsilon}_{\rm gw}\simeq 0.7$,
in the second line.

On the other hand, the peak frequency of the gravitational waves is determined by the Hubble parameter at the decay time.
Let us denote this frequency as $f_h$. We can estimate the present value of the frequency $f_h$ by multiplying the scale $k/a(t_{\rm dec})= 2\pi H(t_{\rm dec})$
by the redshift factor $a(t_{\rm dec})/a(t_0)$,
\begin{equation}
f_h(t_0) = \frac{k}{2\pi a(t_0)} = \frac{a(t_{\rm dec})}{a(t_0)}H(t_{\rm dec}). \label{eq5-8}
\end{equation}
Using the relation
\begin{equation}
\frac{a(t_{\rm dec})}{a(t_0)} = 3.12\times 10^{-30}\times \lambda^{1/4}\epsilon^{-1/2}\left(\frac{\eta}{10^{15}\mathrm{GeV}}\right)^{-1/2}, \nonumber
\end{equation}
we obtain
\begin{equation}
f_h(t_0) \simeq 6.7\times 10^9\times \lambda^{-1/4}\epsilon^{1/2}\left(\frac{\eta}{10^{15}\mathrm{GeV}}\right)^{1/2}\mathrm{Hz}. \label{eq5-9} 
\end{equation}
In Sec.~\ref{sec4}, we observed that the spectrum extends up to the wavenumber $k/a\simeq 2\pi\delta_w^{-1}$, which corresponds to the width of domain walls.
The frequency $f_w$ determined by this length scale can also be estimated as
\begin{equation}
f_w(t_0) = \frac{a(t_{\rm dec})}{a(t_0)}\delta_w^{-1} \simeq 3.4\times 10^9\times\lambda^{3/4}\epsilon^{-1/2}\left(\frac{\eta}{10^{15}\mathrm{GeV}}\right)^{1/2}\mathrm{Hz}. \label{eq5-10}
\end{equation}

Though the peak amplitude and frequency can be determined by the use of Eqs.~\eqref{eq5-7} and~\eqref{eq5-9},
the estimation of the amplitude at high frequencies $f>f_h$ is subtle.
In the previous studies~\cite{Hiramatsu:2010yz,Kawasaki:2011vv}, we extrapolated the numerical results, in which the slope of the spectrum
becomes almost flat in the intermediate frequency range $f_h<f<f_w$.
However, the updated spectrum obtained in Sec.~\ref{sec4} indicates that the amplitude at $f=f_w$ is suppressed
compared with that at $f=f_h$.
It is not so straightforward to estimate the degree of suppression, but the results of the current simulations
indicate the behavior $f^{-1}$ in the intermediate frequencies $f_h<f<f_w$.

We note that the spectrum at low frequencies $f<f_h$ can be deduced from the requirement of causality~\cite{Caprini:2009fx}.
To discuss this point, let us write the spectrum of gravitational waves at the production time $t=t_*$ in the radiation dominated era as~\cite{Kawasaki:2011vv}
\begin{equation}
\Omega_{\rm gw}(k,t_*)\equiv \frac{1}{\rho_c(t_*)}\frac{d\rho_{\rm gw}(t_*)}{d\ln k} = \frac{4}{3\pi^2}k^3\int^{\tau_*}_{\tau_p}\frac{d\tau_1}{\tau_1}\int^{\tau_*}_{\tau_p}\frac{d\tau_2}{\tau_2}
\cos(k(\tau_1-\tau_2))\Pi(k,\tau_1,\tau_2), \label{eq5-11}
\end{equation}
where $\tau_p$ is some onset of the production of gravitational waves.
In the above equation, $\Pi(k,\tau_1,\tau_2)$ represents the unequal time correlator of the anisotropic stress tensor
\begin{align}
\sum_{ij}\langle\Pi_{ij}(\tau_1,{\bf k})\Pi_{ij}^*(\tau_2,{\bf k'})\rangle &\equiv (2\pi)^3\delta^{(3)}({\bf k}-{\bf k'})\Pi(k,\tau_1,\tau_2), \label{eq5-12}\\
a^{-2}(\tau)T^{\rm TT}_{ij}(\tau,{\bf k}) &\equiv (\rho_c+p_c)\Pi_{ij}(\tau,{\bf k}), \label{eq5-13}
\end{align}
where $\langle\dots\rangle$ represents an ensemble average, and $p_c$ is the pressure of the background fluids ($p_c=\rho_c/3$ in the radiation dominated universe). 
Note that $\Pi(k,\tau_1,\tau_2)$ can be written in terms of the Fourier transform of the correlation function in the real space
\begin{align}
\Pi(k,\tau_1,\tau_2) &= \int d^3z e^{i{\bf k\cdot z}}\Pi(z,\tau_1,\tau_2)= \int^{\infty}_0 dz \frac{4\pi z}{k}\sin(kz)\Pi(z,\tau_1,\tau_2), \label{eq5-14} \\
\Pi(|{\bf x}-{\bf x'}|,\tau_1,\tau_2) &\equiv\sum_{ij}\langle\Pi_{ij}(\tau_1,{\bf x})\Pi_{ij}(\tau_2,{\bf x'})\rangle
\qquad {\rm with} \qquad \Pi_{ij}(\tau,{\bf x}) = \int \frac{d^3k}{(2\pi)^3}e^{i{\bf k}\cdot{\bf x}}\Pi_{ij}(\tau,{\bf k}). \label{eq5-15}
\end{align}
From causality it is required that $\Pi(|{\bf x}-{\bf x'}|,\tau_1,\tau_2)=0$ for $|{\bf x}-{\bf x'}| > l_c$,
where $l_c \gtrsim H^{-1}$ is the correlation length. This fact implies that
the integration over $z$ in Eq.~\eqref{eq5-14} is truncated at $z=l_c$.
Therefore, if we consider the mode satisfying $kl_c \ll 1$, the integrand in Eq.~\eqref{eq5-14} can be expanded 
as a power series of $kz$. As a result, $\Pi(k,\tau_1,\tau_2)$ approaches a value which does not depend on $k$ in the limit $kl_c \to 0$,
if the leading contribution of the power series expansion does not vanish.
In this case, the slope of the spectrum is purely determined by the factor $k^3$ in Eq.~\eqref{eq5-11} for small $k$.
This behavior can be checked in part, by computing the equal time correlation function $\Pi(k,\tau,\tau)$
directly from the results of numerical simulations~\cite{Kawasaki:2011vv}.
We compute $\Pi(k,\tau,\tau)$ for the sets of simulations described in Sec.~\ref{sec4} and
observed that $\Pi(k,\tau,\tau)$ hardly depends on $k$ for $k/aH\ll 1$.

\begin{figure}[htbp]
\begin{center}
\includegraphics[scale=1.0]{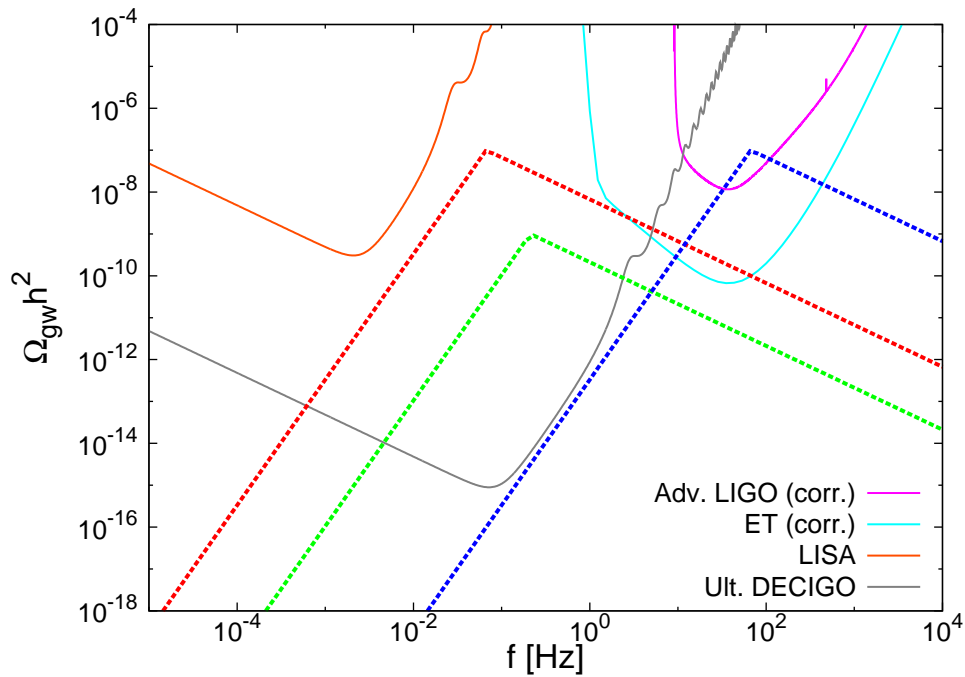}
\end{center}
\caption{The schematics of the gravitational wave signatures from domain walls and sensitivities of planned detectors.
The broken line represents the results estimated by using Eqs.~\eqref{eq5-7} and~\eqref{eq5-9} for the sets of parameters
$(\eta,\epsilon)=(10^{10}\mathrm{GeV},10^{-17})$ (red),
$(\eta,\epsilon)=(10^{10}\mathrm{GeV},10^{-16})$ (green), and $(\eta,\epsilon)=(10^{12}\mathrm{GeV},10^{-13})$ (blue).
Other parameters are fixed such that $\lambda=1$ and $g_*=100$.
While we fix the value of $\lambda$ in this figure, we note that the amplitude depends on $\lambda$ as $\Omega_{\rm gw}\propto \lambda^2$.
The frequency dependence $\Omega_{\rm gw}\propto f^3$ for $f<f_h$ and $\Omega_{\rm gw}\propto f^{-1}$ for $f>f_h$
suggested by the numerical results is also plotted.
The sensitivity curves for LISA and DECIGO are estimated from Eq.~\eqref{eq5-16}, and 
those for LIGO and ET are estimated from Eq.~\eqref{eq5-17}, in which the correlation between two detecters is assumed.
The noise curves $S_n^{1/2}(f)$ are taken from~\cite{ALIGO} for Advanced LIGO and~\cite{ET} for ET, respectively.
For LISA and Ultimate DECIGO, we used the online sensitivity curve generator~\cite{sens} with the parameters taken from Table 7 of Ref.~\cite{Alabidi:2012ex}.}
\label{fig7}
\end{figure}

Figure~\ref{fig7} shows schematics for the spectrum of gravitational waves produced by domain walls in comparison with sensitivities of planned detectors
such as (Advanced) LIGO~\cite{Abramovici:1992ah}, 
ET~\cite{Sathyaprakash:2012jk}, LISA~\cite{AmaroSeoane:2012km}, and Ultimate DECIGO~\cite{Seto:2001qf}.
For LISA and DECIGO, we estimated the sensitivity curve in terms of the formula for the minimum detectable amplitude of $\Omega_{\rm gw}$ in the single detector~\cite{Maggiore:1900zz,Maggiore:1999vm}:
\begin{equation}
\Omega_{\rm gw}^{\rm 1d,min} = \frac{4\pi^2}{3H_0^2}\frac{(\mathrm{SNR})^2}{F}f^3S_n(f), \nonumber
\end{equation}
which reduces to
\begin{equation}
\Omega_{\rm gw}^{\rm 1d,min}h^2 \simeq 0.0125\frac{(\mathrm{SNR})^2}{F}\left(\frac{f}{100\mathrm{Hz}}\right)^3\left(\frac{S_n^{1/2}}{10^{-22}\mathrm{Hz}^{-1/2}}\right)^2, \label{eq5-16}
\end{equation}
where $S_n(f)$ is the noise spectral density, $\mathrm{SNR}$ is the signal-to-noise ratio for a confident detection, and $F$ is the angular efficiency factor of the detector. 
Here we fix the value of $F$ as $2/5$, which corresponds to the interferometer with perpendicular arms~\cite{Maggiore:1900zz,Maggiore:1999vm},
and use a reference value $\mathrm{SNR}=5$.
On the other hand, for LIGO and ET, we assume to take the correlation between two detectors, which improves the sensitivity by several orders of magnitude~\cite{Maggiore:1900zz,Maggiore:1999vm}:
\begin{align}
\Omega_{\rm gw}^{\rm 2d,min}h^2 &\simeq (2T\Delta f)^{-1/2}\Omega_{\rm gw}^{\rm 1d,min}h^2 \nonumber\\
& \simeq 3.98\times 10^{-5}\left(\frac{1\mathrm{year}}{T}\right)^{1/2}\left(\frac{10\mathrm{Hz}}{\Delta f}\right)^{1/2}\Omega_{\rm gw}^{\rm 1d,min}h^2, \label{eq5-17}
\end{align}
where $T$ is the total observation time, and $\Delta f$ is the frequency resolution of the detectors.
In the plots shown in Fig.~\ref{fig7}, we choose $T=1\mathrm{year}$ and $\Delta f = f/10$.
We note that Eq.~\eqref{eq5-17} does not take account of the separation between two detectors.
In the realistic cases, the sensitivity would be suppressed at high frequencies due to the fact that two detectors are placed at different location and have 
a different angular sensitivity. Here we do not include this reduction factor for the sake of a rough comparison.

In Fig.~\ref{fig7},
we plot the peak amplitude and frequency obtained from Eqs.~\eqref{eq5-7} and~\eqref{eq5-9} for
selected values of $\eta$ and $\epsilon$ satisfying the condition~\eqref{eq2-9}.
Note that the peak amplitude and frequency are controlled by three theoretical parameters ($\lambda$, $\epsilon$, and $\eta$)
such that $\Omega_{\rm gw}\propto \lambda^2\epsilon^{-2}\eta^4$ and $f_h \propto \lambda^{-1/4}\epsilon^{1/2}\eta^{1/2}$.
The speculative estimates for the slope of the spectrum ($\Omega_{\rm gw}\propto f^3$ for $f<f_h$ and $\Omega_{\rm gw}\propto f^{-1}$ for $f>f_h$)
are also shown as the broken lines.
As shown in this figure, if the peak frequency corresponding to the horizon scale at the decay time of domain walls lies within
the frequency band in which the planned detectors are sensitive, the signatures of gravitational waves can be probed in future experimental studies.

\section{\label{sec6}Conclusion}
In this paper, we have calculated the spectrum of gravitational waves produced by domain walls based on the lattice simulations
with improved dynamical ranges.
The large dynamical range enables us to survey the dependence on the theoretical parameter $\lambda$,
which controls the energy and width of domain walls. It is found that the spectrum has a peak at the wavenumber
corresponding to the horizon size, regardless of the value of $\lambda$, and that it falls off at a large wavenumber.
The location of the falloff varies with the value of $\lambda$, which indicates that this scale corresponds to the width of domain walls.
The slope of the spectrum in the intermediate scales behaves like $\propto k^{-1}$.
This result is different from the previous one~\cite{Kawasaki:2011vv}, where the spectrum becomes almost flat, but
this difference is caused by the error in the numerical code used in the previous studies.

From the results of the numerical simulations, the scaling parameter $\mathcal{A}$ and the 
differentiation of the efficiency parameter $\tilde{\epsilon}_{\rm gw}$ at the peak of the spectrum
are determined as Eqs.~\eqref{eq4-1} and~\eqref{eq4-2}.
Using the values of $\mathcal{A}$ and $\tilde{\epsilon}_{\rm gw}$ obtained from the simulations, we estimate the amplitude of gravitational waves
observed today as Eq.~\eqref{eq5-7}.
The estimation of $\tilde{\epsilon}_{\rm gw}$ involves large statistical uncertainties, which is caused by 
the variation of the results for different realizations and the uncertainty in the integration over the direction of ${\bf k}$ in $k$-space.
As a result, the estimation for the present density of gravitational waves~\eqref{eq5-7} contains the uncertainty as a factor of $\mathcal{O}(1)$.
The present value of the peak frequency is also estimated as Eq.~\eqref{eq5-9}.
If the scale of the symmetry breaking $\eta$ is sufficiently high and domain walls exist for a long time $\epsilon\ll 1$,
there exists a parameter region where the signature of gravitational waves can be probed in the future gravitational wave interferometers.

\begin{acknowledgments}
We thank B.~S.~Sathyaprakash for notifying about the ET sensitivity curve.
Numerical computation in this work was carried out at the 
Yukawa Institute Computer Facility. 
M.~K.~is supported by Grant-in-Aid for Scientific research 
from the Ministry of Education, Science, Sports, and Culture
(MEXT), Japan, No.~25400248 and No.~21111006
and also by World Premier International Research Center Initiative
(WPI Initiative), MEXT, Japan.
K.~S.~is supported by the Japan Society for the Promotion of Science (JSPS) through research fellowships.
T.~H.~is supported by JSPS Grant-in-Aid for Young Scientists (B) No.~23740186, and also by MEXT HPCI Strategic Program.
\end{acknowledgments}

\appendix
\section{\label{secA}The coding error}
The numerical code used in this paper is developed based on that used in Refs.~\cite{Hiramatsu:2012gg,Hiramatsu:2012sc}.
In this code, we modify some schemes from the old code used in~\cite{Hiramatsu:2010yz,Kawasaki:2011vv}, improving its operation speed.
However, we found that the old code contains an error in evaluating the TT projection of the stress-energy tensor of the scalar field,
which leads to the wrong estimation for the gravitational wave spectrum.

The error is just caused by a technical reason.
We use the discrete
spatial coordinate on the lattice, ({\tt ix}, {\tt iy}, {\tt iz}), where the indices
{\tt ix}, {\tt iy}, {\tt iz} are integers taking a value from {\tt 0} to {\tt N-1}, and the scalar field
defined on the lattice is stored to an 1D array, {\tt f[idx]} as
\begin{align}
\phi({\bf x}) \to {\tt f[idx]} \qquad \mathrm{with} \qquad {\tt idx = (N*iz+iy)*N+ix}. \label{eqA-1}
\end{align}
Also for the each element of stress-energy tensor $T_{ij}$, we store
its data to another 1D array, {\tt T11[idk]}, {\tt T12[idk]}, ..., but we
incorrectly assigned them in the old code as
\begin{align}
T_{ij}({\bf k}) \to {\tt Tij[idk]} \qquad \mathrm{with} \qquad {\tt idk = (N*ix+iy)*N+iz}, \label{eqA-2}
\end{align}
when we evaluate the TT projection defined in Eq.~\eqref{eq3-12}.
Clearly, {\tt idk} should be calculated in the same way as~\eqref{eqA-1}.
As a result, this mismatch leads to a wrong result, since this treatment
corresponds to interchanging $k_x$ and $k_z$ in evaluating Eq.~\eqref{eq3-12}.

\begin{figure}[htbp]
\begin{center}
\includegraphics[scale=0.4]{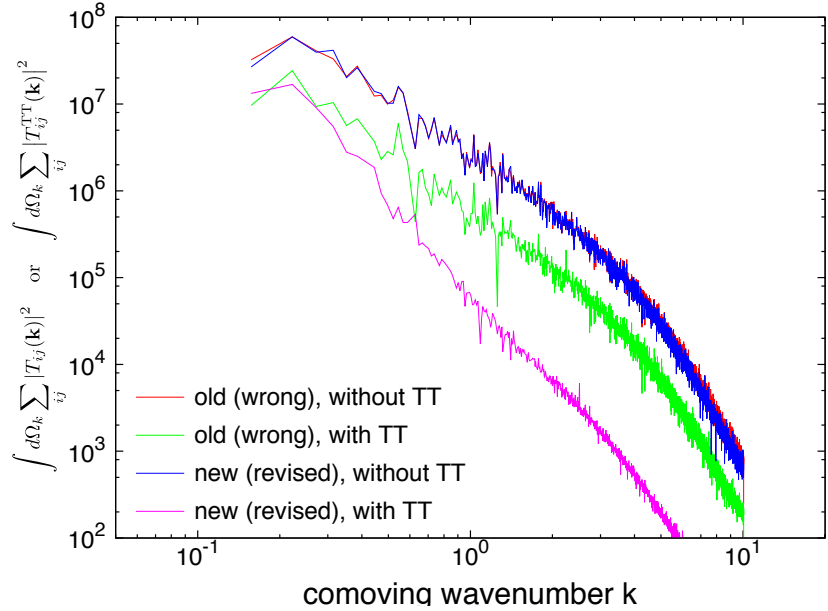}
\end{center}
\caption{The comparison of the results between the old (wrong) code used in Refs.~\cite{Hiramatsu:2010yz,Kawasaki:2011vv}
and the new (revised) code used in this paper.
The vertical axis represents the magnitude of the power of the stress-energy tensor~\eqref{eqA-3} for lines denoted as
``without TT", and that of the TT projection of the stress-energy tensor~\eqref{eqA-4} for lines denoted as ``with TT".
These results are evaluated at $\tau=20$ in the simulation with the grid size $N=128$.
Other parameters are taken as $L=40$ and $\lambda=0.1$.
}
\label{fig8}
\end{figure}

In Fig.~\ref{fig8}, we plot the comparison between the results of the old code used in~\cite{Hiramatsu:2010yz,Kawasaki:2011vv}
and new code used in this paper. Here, we compute the power of the stress-energy tensor
\begin{equation}
\int d\Omega_k \sum_{ij}\left|T_{ij}({\bf k})\right|^2, \label{eqA-3}
\end{equation}
and its TT projection
\begin{equation}
\int d\Omega_k \sum_{ij}\left|T^{\rm TT}_{ij}({\bf k})\right|^2. \label{eqA-4}
\end{equation}
Note that the difference does not appear when we compute the power without TT projection~\eqref{eqA-3}, since it depends only on $|{\bf k}|=\sqrt{k_x^2+k_y^2+k_z^2}$.
The effect of the wrong assignment~\eqref{eqA-2} arises when we compute the power with TT projection~\eqref{eqA-4},
since it depends not only on $|{\bf k}|$, but also on the direction $\hat{k}$.
Accordingly, the old code 
incorrectly estimates the spectrum of gravitational waves.
As shown in Fig.~\ref{fig8}, for the stress-energy tensor of domain walls,
the old code overestimates the power at large wavenumbers,
which leads to the almost flat spectrum observed in the previous numerical studies~\cite{Hiramatsu:2010yz,Kawasaki:2011vv}.



\end{document}